\documentclass[12pt]{iopart}

\usepackage{iopams}
\usepackage{color}
\usepackage{graphicx}
\usepackage{dcolumn}
\usepackage{bm}
\usepackage[utf8]{inputenc}
\usepackage[T1]{fontenc}
\usepackage{mathptmx}
\usepackage{enumerate}
\usepackage{theorem}
\usepackage{xcolor}
\usepackage{caption}
\usepackage{subcaption}
\usepackage{cite}

\newtheorem{thm}{Theorem}

\newtheorem{prp}{Proposition}
\newtheorem{cor}{Corollary}
\newtheorem{defi}{Definition}

\newtheorem{hyp}{Hypothesis}

\newcommand{\proof}{\noindent {\bf Proof:} \hspace{0.1in}}
\newcommand{\qed}{\hfill\mbox{\raggedright $\Box$}\medskip}

\newcommand{\R}{\mathbb{R}}
\newcommand{\Z}{\mathbb{Z}}
\newcommand{\T}{\mathbb{T}}
\newcommand{\M}{\mathsf{M}}
\newcommand{\MS}{\mathsf{S}}
\newcommand{\K}{\mathsf{K}}
\newcommand{\I}{\mathsf{I}}
\newcommand{\metric}{\ensuremath{\mathrm{g}}}
\newcommand{\submetric}{\ensuremath{\mathrm{h}}}
\newcommand{\normal}{\ensuremath{\mathrm{u}}}

\newcommand{\grad}{\raisebox{0.01em}{\scalebox{1.1}[1.4]{${\scriptstyle \nabla}$}}}
\newcommand{\gradbar}{\raisebox{0.01em}{\scalebox{1.1}[1.4]{${\scriptstyle \overline{\nabla}}$}}}

\newcommand{\energy}{\ensuremath{\mathrm{\rho}}}
\newcommand{\energyint}{\ensuremath{\mathrm{\mu}}}

\newcommand{\pressure}{\scalebox{0.9}[1.0]{\ensuremath{\mathsf{P}}}}

\newcommand{\anisotropy}{\scalebox{1.1}[1.0]{\ensuremath{\Sigma}}}

\newcommand{\curvature}{\ensuremath{\overline{\mathrm{R}}}}

\newcommand{\constant}{\scalebox{0.9}[1.1]{\ensuremath{\mathtt{C}}}}

\newcommand{\hubble}{\ensuremath{\mathrm{H}}}
\newcommand{\hubbleA}{\ensuremath{\ensuremath{\sigma_A}}}

\newcommand{\omegap}{\ensuremath{\widehat{\Omega}}}

\newcommand{\shearS}{\ensuremath{\mathsf{S}}}

\newcommand{\submetrict}{\ensuremath{\gamma}}

\begin{document}

\title[Breaking the Cosmological Principle into pieces]{Breaking the Cosmological Principle into pieces: a prelude to the intrinsically homogeneous and isotropic spacetimes}

\author{L. G. Gomes}
\address{Federal University of Itajub\'a (UNIFEI), Av. BPS, 1303, Itajub\'a-MG, 37500-903, Brazil}
\eads{\mailto{lggomes@unifei.edu.br}}

\date{\today}
%
\begin{abstract}
In this manuscript, we show that there are three fundamental building blocks supporting the Cosmological Principle. The first of them states that there is a special frame in the universe where the spatial geometry is intrinsically homogeneous and isotropic. The second demands the existence of a fiducial observer to whom the Hubble parameter is isotropic. The last piece states that matter and radiation behave as a perfect fluid. We show that these three hypotheses give us the Friedmann-Lemaître-Robertson-Walker (FLRW) spacetimes, the central pillar of the standard model of Cosmology. We keep with the first of them and start to investigate the so-called intrinsically homogeneous and isotropic spacetimes. They emerge after the decoupling of the CMB with the geometric frame of reference. Furthermore, a ``$\Lambda$CDM-like" effective theory arises naturally in those backgrounds, together with some new density parameters relating to the local inhomogeneities, the internal energy density, and the local and global magnitudes of the Hubble anisotropy. All those properties make this class of inhomogeneous models, which roughly speaking, keeps "1/3" of the Cosmological Principle, worth investigating in applications to Cosmology, for it can accommodate the same ingredients of the standard model, as a geometric frame and a free-falling isotropic cosmic background radiation, and reduce to the latter when some observable parameters vanish.   
\end{abstract}

\maketitle

\section{\label{sec:Intro} Introduction \protect}

%

Cosmology has been established on the belief that the Universe is homogeneous and isotropic on large scales. Although this is a quite vague notion, it is largely accepted that, when put on sound mathematical grounds, the Cosmological Principle is equivalent to demanding the spacetime $\M$ to be of the FLRW type, named after their precursors Friedmann-Lemaître-Robertson-Walker. It stands as a framework around which the standard model of Cosmology is built \cite{Peebles,Weinberg}. Nonetheless, the recent observational data have been testing the limits of this theory, and their tenets were put under intense scrutiny after the discovery of many ``tensions" \cite{ABDALLA202249}. In particular, the observational basis for the Cosmological principle has been investigated in many aspects \cite{ObservationalCosmologicalPrinciple}, and the doubts on whether it is indeed suitable to describe our large-scale universe have been growing in the last few years\cite{Anisotropy01, Anisotropy02}. Regardless of the outcome of this cosmic conundrum but still inspired by this current debate, in this manuscript, we shall pay a visit to the mathematical hypotheses that lead us to the FLRW spacetimes, keeping those of more fundamental relevance, while re-considering the role of the others, for which the validity is a matter of observational verification.           

The FLRW spacetimes are characterized by their highly symmetric spatial sections: there is a Lie group acting on $\M$ by isometries whose orbits are space-like hypersurfaces, the homogeneity condition, with a transitive induced action of the isotropy group on their tangent spaces, the isotropic character of the action. A well-known theorem ensures the existence of a special coordinate representation where the metric is completely characterized by the scale factor $a(t)$ and the constant $K_0$ representing the spatial curvature \cite{Wald}. As we open up the many restrictions imposed by the large symmetry group of the FLRW spacetimes, we come to the conclusion that such models are characterized by the vanishing of many spatial anisotropies. In fact, the spatial geometry, the Hubble expansion, the energy flux, and the pressure coming from the matter content, impose no spatial asymmetry along the many allowed directions in such spacetimes. 

As we move towards the foundations of the Cosmological Principle, we first realize that it demands space to have the simplest of the geometries, those which do not distinguish two different directions nor distinct points. These homogeneous and isotropic environments are the simply connected Riemannian manifolds with constant sectional curvature $K_0$, which are restricted to three types of space: The Euclidean ($K_0=0$), the hyperbolic ($K_0<0$), and the spheric ($K_0>0$)  \cite{Wolf, vinberg}. We can set this condition as the existence of special observers in the Universe, represented by a time-like unitary vector field $\normal$, to whom the notions of space and time are naturally distinguished, that is, $\normal$ has vanishing vorticity \footnote{If $\normal$ has non-vanishing vorticity, then its orthogonal space distribution is no longer integrable (\cite{ellis_mac_marteens}, chapter 4). Hence, the notion of ``space" to those observers cannot be given by a unique natural choice.}, such that the equal-time spatial sections have one of those simplest configurations. They are formally defined in hypothesis H\ref{Hyp:IsotropyCurvature}, and the resulting structure is called intrinsically homogeneous and isotropic spacetime, the subject of investigation in this manuscript. They constitute a subclass of the broader family of spacetimes admitting intrinsic symmetries \cite{coll_79, coll1_79, coll2_79}.

The second fundamental notion behind the Cosmological Principle is the isotropy of the Hubble sky, that is, the property of expanding equally in all directions. It has a strong observational appeal, so much so that its validity has been questioned in some recent data analyses (see Ref.\cite{ObservationalCosmologicalPrinciple} and references therein). Regardless of the constraints imposed by observational Cosmology, this piece of the Cosmological Principle has its own subtleties to be understood and dealt with even on a theoretical basis. Nonetheless, the Hubble isotropy dwells in the core of the standard model of Cosmology and is largely accepted in the scientific community. The intrinsically homogeneous and isotropic spacetimes readily recover this feature under the assumption of the existence of a fiducial observer to whom the Hubble sky is the same in all directions. In addition, we also assume that it is able to distinguish a scale factor $\ell$ such that the local volume grows with $\ell^{3}$, if we are in a four-dimensional spacetime, and the curvature decreases as $\ell^{-2}$. This is the second piece of the Cosmological Principle, written down in hypothesis H\ref{Hyp:IsotropyHubble}. 

The third and last piece is the isotropy of the physical constituents of matter and radiation. It has a strong physical appeal by stating that there is no net energy flux nor anisotropic stress in the universe. In fact, as we settle the cosmological observers as those to whom the cosmic background radiation (CMB) is nearly isotropic, for instance, it represents the idea that there would be no giant magnetic field, nor any constituent of matter moving in a relativistic speed, nor any phenomena that could account for significant breaking of the perfect fluid form of the energy-momentum tensor. It is set in the hypothesis H\ref{Hyp:IsotropyEMTensor}.  

Those are the three building blocks of the Cosmological Principle, put down in section\ref{Sec:CosmologicalPrinciple}. The first, mathematical, ensures the homogeneity and isotropy of the spatial geometry to describe the universe. The second, observational, guarantees isotropic expansion. The third, physical, assumes a perfect fluid form for the overall combination of matter and radiation. As we put all of them together and assume Einstein's equations to hold, we recover the FLRW spacetime, with each one of these assumptions being absolutely necessary for it. We discuss that in section \ref{Sec:H1+H2+H3+EEq}.

The main purpose of this manuscript is to start investigating the intrinsically homogeneous and isotropic spacetimes, whose mathematical foundations are the subject of section \ref{Sec:IntrinsicallyHomogeneousIsotropicSpacetimes}. This is a framework where just the first ``1/3" of the Cosmological Principle is kept. It turns out that this structure naturally appears when we allow the CMB observers to decouple from these geometric ones, as discussed in section \ref{Sec:ReferenceFrame}. It corresponds to that counterpart less accessible by our direct observations or our physical intuition, and therefore, less likely to be ruled out by physical considerations or surveys probing the cosmological sky. In this sense, the hypothesis of homogeneity and isotropy of spatial geometry seems to be the one to keep with, while leaving the rest "2/3" to observational scrutiny. Even large-scale homogeneity and a $\Lambda$CDM-like approach to Cosmology naturally arise from this rich framework, as we show in sections \ref{Sec:PeriodicConditions} and \ref{Sec:EffectiveModel}, respectively. The notations and sign conventions follow Ref. \cite{ellis_mac_marteens}.

\section{On the hypotheses laying in the foundations of the Cosmological Principle}
\label{Sec:CosmologicalPrinciple}

Here we follow a methodology closely related to the co-moving approach of General Relativity, very often used in Cosmology \cite{Hwk_Ellis, ellis_mac_marteens}, differing from it only in some choices of variables and in the interpretation that our reference system will not be necessarily co-moving with a fluid. We start with a $m$-dimensional spacetime $\M$ endowed with the Lorentzian metric $\metric$ and establish our cosmological observers by choosing a time-like vector field $\normal$, oriented to the future, unitary, $\normal^{\, 2}=-1$, and vorticity-free, that is, its orthogonal distribution $\normal^\bot$ is integrable. This last condition ensures that we are not breaking the isotropy condition from the beginning, for otherwise, we would have the vorticity vector pointing to some preferred direction, at least if $\dim \M=4$. It stands as the intuitive notion of a ``rigid" system of observables, with no relative rotation among its constituents. Furthermore, the curvature of the spatial sections orthogonal to $\normal$ should not distinguish between two different points or directions, meaning that our space sections have constant sectional curvature \cite{Wolf, vinberg}. 

Looking for a conceptual understanding of the mathematical foundations behind the Cosmological Principle, we set our first hypothesis as follows:
\begin{hyp}\label{Hyp:IsotropyCurvature}
In the spacetime $\M$ with metric $\metric$ there is a time-like vector field $\normal$, oriented to the future, unitary, $\normal^{\, 2}=-1$, and vorticity-free. For each point $p \in \M$, the maximum integrable manifold  of $\normal^\bot$ passing through $p$, $\MS_p$, the spatial section at $p$, is a space of constant curvature $K(p)$ with the inherited Riemannian metric
\begin{equation}\label{Eq:SpaceMetric}
\submetric = \metric + \normal^\flat\normal^\flat \, , 
\end{equation}
where $\normal^\flat := \metric(\normal, \cdot )$ is the $1$-form naturally associated to $\normal$ through $\metric$, the lowering of its index. Any two spatial sections are diffeomorphic. Such spacetimes are called intrinsically homogeneous and isotropic.  
\end{hyp}
Throughout the manuscript, we assume hypothesis \ref{Hyp:IsotropyCurvature} to hold, so that our investigation is concentrated on the general intrinsically homogeneous and isotropic spacetimes. In this context, there is a natural splitting in our spacetime settling our choices for time and space, respectively, notions of which can be incorporated into any tensor field. For instance, we will denote $\varphi: \M \to \R$ by a time function whenever $d\varphi(p) \cdot v = 0$ for every $p \in \M$ and $v \in \normal(p)^\bot$. Analogously, it is a space function if $d\varphi \cdot \normal = 0$. A vector field $X$ is a time vector field if it is everywhere proportional to $\normal$ and a space vector field if it is everywhere orthogonal to $\normal$. Any covariant tensor field in $\M$ is also referred to as a space tensor field if it vanishes whenever contracted with $\normal$, and so on. Moreover, the spatial gradient is the spatial projection of the gradient, which coincides with the gradient defined in the intrinsic Riemannian geometry of each spatial section (denote $\normal \cdot \phi := \normal^\nu\partial_\nu\phi$) :
\begin{equation}\label{Eq:SpatialGrad}
\gradbar \phi = \grad\phi + (\normal \cdot \phi) \, \normal \, .
\end{equation}

An important role is played by the acceleration vector field $\grad_\normal\normal$. Besides being orthogonal to $\normal$, its associated one form $\grad_\normal\normal^\flat := \metric(\grad_\normal\normal, \cdot )$, which in coordinate representation is simply $\grad_\normal\normal_\nu$, is closed when restricted to each space section $\MS_p$, a fact that is equivalent to the vanishing of the vorticity of $\normal$ (Compare with formula (4.36) in \cite{ellis_mac_marteens}). From that intrinsic viewpoint, it is locally the spatial gradient of a ``potential". In a more precise manner, in the following lemma, we establish this well-known result, for the sake of completeness: 
\begin{prp}\label{Thm:AccelerationGrad}
Let $\normal$ be a time-like and vorticity-free vector field and $U$ an open set in $\M$ such that $\MS_p \cap U$ is connected and simply connected along each spatial section $\MS_p$, $p \in U$. There are smooth functions, a time $t:U \to \R$ and a ``potential" $\phi: U \to \R$, such that each space section $\MS_p$ is locally characterized by $t=t(p)$, that is, $\MS_p \cap U = t^{-1}(t(p))$, and the spatial gradient of $\phi$ gives the acceleration vector, 
\begin{equation}\label{Eq:DefTimePotential}
\gradbar \phi = \grad_\normal\normal \, .
\end{equation}
Furthermore, any two ``potentials" in $U$ differ by a time function in $U$:
\begin{equation}\label{Eq:DefTimePotentialGauge}
\phi_2 - \phi_1 = \varphi(t) \, . 
\end{equation}
\end{prp}
\proof 
With no loss of generality, put $U=\M$ connected and simply connected, and set a curve of reference points in each space section by starting from a point $p_0 \in \M$ flowing along $\normal$, $c(t)=F_\normal^t(p_0)$, $F_\normal^t$ being its flux. If we assume connected space sections, each $p \in \M$ would correspond to a time $t_p$ for which $p \in \Sigma_{c(t_p)}$. Now we take the integral of $\left(\grad_\normal\normal\right)^\flat$ along a path lying entirely in $\Sigma_{c(t_p)}$, from $c(t_p)$ to $p$, 
\begin{equation}\label{Eq:DefTimePotential_Integral}
 \phi (p) = \int_{c(t_p)}^p \,\left(\grad_\normal\normal\right)^\flat \, .
\end{equation}
This defines a function as long as the path in the integral is taken to lie entirely in a simply connected region of $\Sigma_p$, since $\left(\grad_\normal\normal\right)^\flat$ is closed along this space section. By definition, it is clear that the space component of $\grad \phi$ is the acceleration vector field. Assuming that both $\phi_1$ and $\phi_2$ satisfy equation (\ref{Eq:DefTimePotential}), uniqueness follows from the fact that for $\varphi = \phi_2-\phi_1$ and $X$ a space vector field, we have $\grad\phi_i=\lambda_i \normal+\grad_\normal\normal$ and
\[
 d\varphi \cdot X = \metric(\grad \varphi , X) = (\lambda_2-\lambda_1)\, \metric(\normal , X) =0 \, .
\]
\qed

In local coordinates $(x^\mu)=(t,x^i)$ adapted to the observers $\normal$, guaranteed by the vanishing of the vorticity, any  ``potential" $\phi$ appears as $\normal=e^{-\phi}\frac{\partial}{\partial t}$ and in the metric as $\ln \sqrt{-\metric_{00}}$, that is,   
\begin{equation}\label{Eq:MetricGeneral}
 \metric = - \, e^{2 \phi(t,x)}\, dt^2 +  \submetric_{ij}(t,x) \, dx^i dx^j \, .
\end{equation}
Note that $\metric_{00} \approx -(1 + 2\phi)$ in the nearly Newtonian regime \cite{MTW}. Therefore, any such function has a simple interpretation: it is the gravitational potential in the usual textbook-like Newtonian limit of General Relativity \cite{MTW, LGGomes_2023}. Needless to say, this special status of a true potential loses its meaning in General Relativity. Throughout the text, our coordinate system is assumed to be adapted to $\normal$, just like in the formula (\ref{Eq:MetricGeneral}) above.

An $\normal$-observer is a curve $c(t)$ which is an integral line of $\normal$, that is, $\dot{c}(t)=\normal(c(t))$. Due to the gauge freedom of proposition \ref{Thm:AccelerationGrad}, we can always assume $\phi$ to satisfy $\phi(c(t))=0$, so that in $\normal$-adapted coordinates, $c(t)=(t,x_0)$ for a fixed $x_0$. Moreover, $\grad_{\dot{c}(t)}\dot{c}(t)=\grad_{\normal}\normal (c(t))=\gradbar\phi(c(t))$, that is, $c(t)$ is a geodesic if, and only if, $\gradbar\phi(c(t))=0$. This is the relativistic way to express that there is no net gravitational force acting on this observer.

Another important geometric structure induced by $\normal$ is the expansion tensor \cite{ellis_mac_marteens}, $\theta^\mu_\nu$, which is the symmetric component of the spatial projection of $\grad^\mu\normal_\nu$. Since its skew-symmetric counterpart, the vorticity, vanishes, it satisfies  $\theta_{\mu 0}=\theta_{0\mu}=0$, and for the spatial components, $\theta^i_{k}=\grad^i\normal_k$, 
\begin{equation}\label{Eq:DefExpansionTensor}
\theta^i_{k} =\frac{e^{-\phi}}{2}\submetric^{i\ell}\frac{\partial}{\partial t}\submetric_{\ell k} = \hubble\,\left(  \delta^i_{k} + \constant\anisotropy^i_{k}\right) 
\quad , \quad  \anisotropy_k^k=0 \, ,    
\end{equation}
with $\hubble$ the Hubble parameter and $\anisotropy^i_{k}$ the Hubble anisotropy tensor, measuring the mean expansion rate of the spatial length and the anisotropies in it, respectively. The constant in the formula (\ref{Eq:DefExpansionTensor}) appears in order to set $\anisotropy =1$ as a turning point in the generalized Friedmann equation (\ref{Eq:FriedmannEquation}), that is,
\begin{equation}
\constant = \sqrt{(m-1)(m-2)} \quad, \quad m=\dim \M \, .
\end{equation}
The reason to opt for the dimensionless Hubble anisotropy $\anisotropy^i_{k}$ instead of the usual shear $\constant \hubble \anisotropy^i_k$ is justified by its straightforward representation in the Kasner plane and natural interpretation of their magnitudes as regular ($\anisotropy < 1/2$), intermediate ($1/2 < \anisotropy < 1$) and extreme ($\anisotropy > 1$) \cite{LGGomes_2021_IJMPD}, where the Hubble anisotropy magnitude is 
\begin{equation}
\anisotropy = \sqrt{\anisotropy_i^k\anisotropy_k^i} \, .
\end{equation}
At each point $p \in \M$, $\hubble(p)$ is the value of the expansion rate averaged over all possible spatial directions $\hat{\bf n}$ at $p$, $\hubble(p)=\langle \theta_{ik}(p)n^in^k\rangle_{S^{m-2}}$, while $\anisotropy$ is proportional to the standard deviation of the Hubble anisotropy at $p$, $\sigma_A(p)^2=\left\langle \left(\theta_{ik}(p)n^in^k\right)^2\right\rangle_{S^{m-2}} - \hubble(p)^2$.\footnote{If we represent $\hat{\bf n}=n^ie_i(p)$ in an orthonormal basis of $T_p\MS_p$ diagonalizing $\anisotropy^i_k(p)$, whose eigenvalues are $\anisotropy_1, \ldots , \anisotropy_{m-1}$, we have $(n^1)^2+ \ldots + (n^{m-1})^2=1$, $\anisotropy_1 + \ldots + \anisotropy_{m-1}=0$, and $\anisotropy^2=(\anisotropy_1)^2 + \ldots + (\anisotropy_{m-1})^2$. By the symmetries of the sphere, we also have $\langle (n^i)^2 \rangle_{S^{m-2}}=1/(m-1)$, $\langle (n^i)^2 (n^k)^2 \rangle_{S^{m-2}}$ equal to $A_m$, for $i=k$, and $B_m$, for $i \ne k$. Applying these relations, we get $\sigma_A^2/\hubble^2 =  \constant^2 \left\langle \left(\anisotropy_{ik}n^in^k\right)^2\right\rangle_{S^{m-2}} = (m-1)(m-2) (A_m-B_m)\anisotropy^2$. For $m=4$, $A_4=1/5$ and $B_4=1/15$.} For instance,  
\begin{equation}\label{Eq:AnisotropyStandardDeviation}
m=4 \quad \Rightarrow \quad \anisotropy = \frac{\sqrt{5}}{2}\frac{\sigma_A}{|\hubble|} \, .
\end{equation}

The Hubble parameter has another well-known interpretation: the $\normal$-observer passing at $p=(t_1,x_1)$, which in the adapted coordinates (\ref{Eq:MetricGeneral}) is just the curve $c(t)=(t,x_1)$, measures the infinitesimal spatial volume $dV(t)$ at each instant of time, giving rise to the volumetric scale factor at $p$, $\ell_V(t)$, with
\begin{equation}
dV(t)=\ell_V(t)^{m-1} dV(t_1)  \quad \textrm{and} \quad 
\hubble(t,x_1) = \frac{1}{\ell_V(t)}\frac{d\ell_V(t)}{d\tau_1} \, ,
\end{equation}
where $d\tau_1=e^{\phi(t,x_1)}dt$ is its proper-time. This means that $\hubble$ measures the expansion rate of the universe with respect to the volumetric scale factor. In a similar manner, if the scalar curvature of the spatial sections does not vanish along the observer passing at $p$, $\curvature(t,x_1)\ne 0$, we can define the curvature scale factor and its ratio by, respectively, 
\begin{equation}
\curvature(t,x_1)=\frac{\curvature(t_1,x_1)}{\ell_K(t)^{2}}  \quad \textrm{and} \quad 
\hubble_K(t,x_1) = \frac{1}{\ell_K(t)}\frac{d\ell_K(t)}{d\tau_1} \, .
\end{equation}
Our general intuition involving volume, area, and curvature would demand the factors $\ell^{m-1}$, $\ell^{2}$, and $\ell^{-2}$, respectively, as we re-define the length scale by a factor $\ell$. In that case, we would have $\ell_V=\ell_K$. In General Relativity, things can get much more complicated, in part due to the freedom of choice of the observers $\normal$, and consequently to the evolution of the induced spatial geometry. Nonetheless, in the late-time universe, we could imagine that some special observers would be free-falling in a "calm" region of the universe, where our naive basic intuition on re-scaling parameters holds true. For this reason, we define:
\begin{defi}\label{Def:FiducialObservers}
An $\normal$-observer $c(t)$ is called fiducial if the following conditions hold:
\begin{enumerate}
\item It is free-falling, that is, $c(t)$ is a geodesic;   
%
%
\item If $\curvature\ne 0$, then the curvature and volumetric scale factors coincide along $c(t)$;   
\end{enumerate}
\end{defi}

After setting our considerations on the Hubble parameter and its anisotropy, we are ready to formulate what we mean by the second piece of the Cosmological Principle:
\begin{hyp}\label{Hyp:IsotropyHubble}
There is a fiducial $\normal$-observer $c(t)$ to whom the Hubble parameter is isotropic, that is, $\anisotropy(c(t))=0$.
\end{hyp}

The splitting of space and time imposed by the $\normal$-observers naturally and uniquely decomposes the energy-momentum tensor as \cite{ellis_mac_marteens} 
\begin{equation}
T_{\mu\nu} = \energy \normal_\mu\normal_\nu + q_\mu\normal_\nu+ q_\nu\normal_\mu + \pressure \submetric_{\mu\nu} + \pi_{\mu\nu} \, ,
\end{equation}
where $\energy=T_{\mu\nu}\normal^\mu\normal^\nu$ is the energy density, $q_\mu =-\, \left( T_{\mu\nu}\normal^\nu + \energy \normal_\mu\right)$ is the spatial vector field ($q_\mu\normal^\mu=0$) representing the energy flux, $\pressure = (T^\mu_\mu + \energy)/(m-1)$ is the relativistic pressure, and $\pi_{\mu\nu}$ is the anisotropic stress, which is symmetric, spatial ($\pi^\mu_\nu\normal^\nu=0$), and traceless ($\pi^\mu_\mu=0$). We recall that this decomposition does not presuppose a single fluid in our spacetime. In an environment with  many fluids,  $T_{\mu\nu} = T_{\mu\nu}^{(a)}+T_{\mu\nu}^{(b)}+\ldots$ and the energy density is $\energy = \energy^{(a)} + \energy^{(b)} + \ldots$, where $\energy^{(a)} = T_{\mu\nu}^{(a)}\normal^\mu\normal^\nu$, and so on. Hence, the physical parameters of the energy-momentum tensor described above are effective in the sense that they sum all the contributions coming from every component of matter and radiation present in our model. Furthermore, $\normal$ is not necessarily assumed to be co-moving with any of them. 

The last ingredient of the Cosmological Principle is the absence of anisotropies in the energy and momentum as it is seen by the $\normal$-observers:
\begin{hyp}\label{Hyp:IsotropyEMTensor}
The total energy-momentum tensor has a perfect fluid form $T_{\mu\nu} = \energy \normal_\mu\normal_\nu + \pressure \submetric_{\mu\nu}$ in the referential frame of the $\normal$-observers.
\end{hyp}

These three hypotheses put together with the Einstein's equations give us the Cosmological Principle (Theorem \ref{Thm:H1+H2+H3}). The first of them, H\ref{Hyp:IsotropyCurvature}, states that there are special observers in the universe to whom the spatial geometry is intrinsically homogeneous and isotropic, a matter of mathematical aesthetics and simplicity while establishing the framework where Cosmology is built onto. The second, H\ref{Hyp:IsotropyHubble}, has a stronger observational appeal, as it states that there is a preferred observer to whom the Hubble parameter is isotropic in a region where the curvature and the volume scale properly. The third, H\ref{Hyp:IsotropyEMTensor}, concerns the physical aspects of the universe by demanding matter and radiation to behave as an effective perfect fluid. Each one is necessary to obtain the Friedmann-Lemaître-Robertson-Walker (FLRW) spacetime, and therefore, to set the central pillar of the standard model of Cosmology, as we will see in section \ref{Sec:H1+H2+H3+EEq}. In other words, they are the fundamental building blocks of the Cosmological Principle.

\section{Intrinsically homogeneous and isotropic spacetimes}
\label{Sec:IntrinsicallyHomogeneousIsotropicSpacetimes}

\subsection{Local coordinate representation}\label{Sec:LocalCoordinates}

We begin by analyzing the local form of the intrinsically homogeneous and isotropic spacetimes. We recall that among the complete and simply connected Riemannian spaces of constant curvature, there are only three possibilities: the Euclidean space $\mathbb{E}^{m-1}$, the Sphere $\mathbb{S}^{m-1}$ and the Hyperbolic Space $\mathbb{H}^{m-1}$. Any other, as far as it is complete, is the quotient of one of them with a discrete subgroup of isometries. Despite the different topologies they can have, thus differing globally, locally they are all isometric as far as they have the same curvature, $K$, putting them in three distinct homothetic classes, depending on whether $K<0$, $K=0$ or $K>0$ \cite{Wolf,vinberg}. In our general relativistic context, each leaf of the foliation defined along the hypothesis H\ref{Hyp:IsotropyCurvature} behaves just as equal. The difference between the theory of the spaces of constant curvature and our relativistic framework is that we have to cope with the geometry varying with time. In order to avoid topological problems appearing in the aforementioned global context, we assume the curvature to keep its sign along $\M$, that is, $K=0$ throughout $\M$ or $K(p)K(q) > 0$ for any two points $p,q \in \M$. The result is laid down in the following theorem:
\begin{thm}\label{Thm:LocalFormConstantCurvature}
Let $(\M,\metric)$ be an intrinsically homogeneous and isotropic spacetime with respect to the set of observers described by the time-like and vorticity-free vector field $\normal$, $\normal^2=-1$. For each spatial section orthogonal to $\normal$, denote its spatially constant curvature by the time function $K$, which does not change sign along $\M$. Hence, around any point $p \in \M$, there is a chart $\varphi=(t,x)$ defining coordinates where the metric is represented as
\begin{equation}\label{Eq:MetricGeneralConstantCurvature}
 \metric = - \, e^{2 \phi(t,x)}\, dt^2 + 
 \frac{a(t)^2 \submetrict_{ik}(t) \, dx^i dx^k}{\left( \, 1 + \frac{K(t)a(t)^2}{4} \, \submetrict_{ik}(t)(x^i-x^i_0(t))(x^k-x^k_0(t))\, \right)^2}\, \, ,
\end{equation}
with $\det(\submetrict_{ij}(t))=1$, $\varphi(p)=(t_0,0)$, and $x^i_0(t_0)=0$. We call $\varphi$ a canonical coordinate system centered at $p$.
\end{thm}
\proof
The case $K=0$ has been done in Ref.\cite{LGGomes_2022_CQG_2}. According to the hypothesis that $K$ does not change sign, we will assume that $K \ne 0$ for any point in $\M$. Since $\normal$ is vorticity-free, we can take an open neighborhood $\widetilde{U} \subset \M$ of a given point $p_0$ in $\M$ where there is an adapted coordinate system $(\tilde{t},\tilde{x})$, as in formula (\ref{Eq:MetricGeneral}), for which $\widetilde{\submetric}_{ik}(\tilde{t},\tilde{x})$ has constant curvature $K(\tilde{t})$. For each time $\tilde{t}$, since spaces of constant curvature are conformally flat, we can define a function $\psi(\tilde{t},\tilde{x})$ such that $\psi(\tilde{t},\tilde{x})^2\widetilde{\submetric}_{ik}(\tilde{t},\tilde{x})$ is a family of matrices with unitary determinant representing flat metrics parametrized by $\tilde{t}$. In this case, $\psi(\tilde{t},\tilde{x})$ is defined by $\det (\widetilde{\submetric}_{ik}(\tilde{t},\tilde{x}))^{-1/2(m-1)}$, that is, it is smooth. Hence, $\psi^{2}\metric$ is a space-flat metric in $\widetilde{U}$, implying that we can use the proposition 1 of Ref.\cite{LGGomes_2022_CQG_2} in order to obtain a new coordinate system $\varphi=(t,x)$ around $p$, also adapted to the observer $\normal$, such that 
\begin{equation}\label{Eq:ConformalFactorConstantCurvature}
\submetric_{ik}(t,x) = \frac{1}{\psi(t,x)^2}\submetrict_{ik}(t)\, . 
\end{equation}
As we fix $t_0$ with $p$ in the spatial section $t=t_0$ and apply a well-known theorem \cite{Wolf} for this Riemannian manifold with metric $\submetric_{ij}(t_0,x)$ of constant curvature $K_0=K(t_0)$, we can assume, without loss of generality, that the spatial coordinates $x^i$ are centered at $p$ with 
\begin{equation}\label{Eq:InitialPsi}
\varphi(p)=(t_0,0) \quad, \quad \submetrict_{ik}(t_0)=\delta_{ik} 
\quad, \quad \textrm{and} \quad  
\psi(t_0,x)=1 + (K_0/4)\delta_{ik}x^ix^k \, .    
\end{equation} 
Since the curvature tensor of $\submetric_{ik}(t,x)$, as a metric with constant curvature, is
\[
\curvature_{ijk\ell} = K(t)\psi(t,x)^{-4}\left(\submetrict_{ik}(t)\submetrict_{j\ell}(t) - \submetrict_{i\ell}(t)\submetrict_{jk}(t)\right)\, , 
\]
a straightforward calculation of $\curvature_{ijk\ell}$ using the relation (\ref{Eq:ConformalFactorConstantCurvature}) and posterior comparison with the formula above lead us to
\begin{equation}
\chi_{ik}\submetrict_{j\ell}(t) +\chi_{j\ell}\submetrict_{ik}(t) -
\chi_{i\ell}\submetrict_{jk}(t) -\chi_{jk}\submetrict_{i\ell}(t) 
=0 \, ,
\end{equation}
where
\begin{equation}
\chi_{ik} = \left(K(t) - \submetrict^{j\ell}(t)\partial_j\psi \partial_\ell\psi \right) \submetrict_{ik}(t) - 2\psi \partial_i\partial_k\psi  \, .
\end{equation}
If we first contract the above expression with $\submetric^{j\ell}(t)$, we get $(m-3)\chi_{ik}=-\chi \, \submetrict_{ik}(t)$ , $\chi=\chi_{j\ell}\submetrict^{j\ell}(t)$, and then with $\submetrict^{ik}(t)$, we obtain that $\chi=0$, and therefore that $\chi_{ik} = 0$.

For each $t$, define $y^k = A^k_i(t)x^i$, where  $\submetrict_{j\ell}(t)A^j_i(t)A^\ell_k(t)=\delta_{ik}$. The condition $\chi_{ik}=0$ for $i\ne k$ implies $\partial^2\psi/\partial y^i\partial y^k =0$, that is, $\psi = \psi_1(t,y^1)+\ldots + \psi_{m-1}(t,y^{m-1})$. Denoting the derivative with respect to $y^k$ by $\psi_k'$, the condition $\chi_{ik}=0$ implies
\[
\psi_k''(t,y^k) =  \frac{K(t) - [\psi_1'(t,y^1)]^2 - \ldots - [\psi_{m-1}'(t,y^{m-1})]^2}{2 \left( \psi_1(t,y^1)+\ldots + \psi_{m-1}(t,y^{m-1}) \right)} = \frac{A(t)}{2} \, .
\]
This over-determined system can be solved only for  
\[
\psi_k(t,y^k) =  \frac{A(t)}{4}(y^k-y^k_0(t))^2 + b_k(t) \, ,
\]
with $A(t), y^k_0(t),b_k(t)$ smooth functions and $A(t)B(t)=K(t)$ for $B=\sum b_k$. This implies that $\psi(t,y) =  A(t)\delta_{ik}(y^i-y^i_0(t))(y^k-y^k_0(t))/4 + B(t)$. Hence, returning to the $x$-coordinates, we obtain
\[
\psi(t,x) =  \frac{A(t)}{4}\submetrict_{ik}(t)(x^i-x^i_0(t))(x^k-x^k_0(t)) + B(t) \, .
\]
A simple comparison with the initial condition for $\psi$ in equation (\ref{Eq:InitialPsi}) tells us that $A(t_0)=K_0$,  $x_0^k(t_0)=0$ and $B(t_0)=1$. Hence, we set $a(t) = \det(\submetrict_{ij}(t))^{1/2(m-1)}/B(t)$ and change $\submetrict_{ij}(t) = B(t)^2 a(t)^2\widehat{\submetrict}_{ij}(t)$, with $\det(\widehat{\submetrict}_{ij})=1$. As we ommit the hat in $\widehat{\submetrict}_{ij}$, we get the representation given in the formula (\ref{Eq:MetricGeneralConstantCurvature}), thus proving the theorem.
\qed

The formula (\ref{Eq:MetricGeneralConstantCurvature}) is our reference point to investigate the local properties of the intrinsically homogeneous and isotropic spacetimes. Note that in those coordinates, the Hubble parameter and anisotropy tensor of the $\normal$-observers look like as, respectively,
\begin{equation}\label{Eq:FormsHubble+Anisotropy}
\hubble =  e^{-\phi} \left( \hubble_S - \hubble_\psi \right) 
\quad \textrm{and}\quad 
\anisotropy^i_k = \frac{\shearS^i_k}{\constant \left(\hubble_S - \hubble_\psi\right)} \, ,
\end{equation}
where we have defined
\begin{equation}\label{Eq:HubbleFunctions}
\hubble_S = \frac{1}{a}\frac{d a}{d t}
\quad , \quad 
\hubble_\psi = \frac{1}{\psi}\frac{\partial \psi}{\partial t}
\quad \textrm{and} \quad 
\shearS^i_k =\frac{1}{2} \submetrict^{i\ell}  \frac{d }{d t}\submetrict_{\ell k} \, ,
\end{equation}
with $\shearS^k_k=0$, and
\begin{equation}\label{Eq:DefPsi}
\psi = 1+ \frac{K(t)a(t)^2}{4}\submetrict_{ik}(t)(x^i-x^i_0(t))(x^k-x^k_0(t))  \, .
\end{equation}
Hence, we get:
%
\begin{cor}\label{Thm:H1plusH2}
An intrinsically homogeneous and isotropic spacetime is shear-free ($\anisotropy=0$) if, and only if, we can set $\submetrict_{ik}=\delta_{ik}$ in the local representation (\ref{Eq:MetricGeneralConstantCurvature}): 
\begin{equation}\label{Eq:MetricGeneralConstantCurvatureH1plusH2}
 \metric = - \, e^{2 \phi(t,x)}\, dt^2 + 
 \frac{a(t)^2\, \left( (dx^{1})^2 + \ldots + (dx^{m-1})^2 \right)}{\left( \, 1 + \frac{K(t)a(t)^2}{4} 
 \, \left((x^1-x^1_0(t))^2 + \ldots + (x^{m-1}-x^{m-1}_0(t))^2 \right)\, \right)^2} \, .
\end{equation}
\end{cor}
\proof
It is clear that the representation (\ref{Eq:MetricGeneralConstantCurvatureH1plusH2}) follows from (\ref{Eq:MetricGeneralConstantCurvature}) if, and only if, $S_{ik} \sim \dot{\gamma}_{ik}=0$.  
\qed

Formulas as (\ref{Eq:MetricGeneralConstantCurvatureH1plusH2}) are usual to the so-called Stephani-Barnes spacetimes \cite{1973_GRG_Barnes, CMP_1967_Stephani, 1981_GRG_Krasinski, 1983_GRG_Krasinski, Book_Krasinski_1997}, that is, those Hubble-isotropic ($\anisotropy=0$), or shear-free,  where H\ref{Hyp:IsotropyCurvature} and H\ref{Hyp:IsotropyEMTensor} also hold. In that case, the curve $x_0(t)$ is often called the ``wandering center of symmetry". We can eliminate it in the vicinity of the free-falling observers to whom the Hubble parameter looks homogeneous in the first-order approximation. Rigorously, we have: 
\begin{thm}\label{Thm:WanderingCenterSymmetry}
In an intrinsically homogeneous and isotropic spacetime where the curvature does not change sign, around each free-falling $\normal$-observer $c(t)$ for which the Hubble parameter is spatially homogeneous up to the first order, that is, $\gradbar\hubble (c(t))=0$, there is a coordinate system adapted to $\normal$ where $c(t)=(t,0)$ and the metric is
\begin{equation}\label{Eq:MetricGeneralConstantCurvatureFiducial}
 \metric = - \, e^{2 \phi(t,x)}\, dt^2 + 
 \frac{a(t)^2 \submetrict_{ik}(t) \, dx^i dx^k}{\left( \, 1 + \frac{K(t)a(t)^2}{4} \, \submetrict_{ik}(t)x^ix^k\, \right)^2}\, \, ,
\end{equation}
with $\det(\submetrict_{ij}(t))=1$. In particular, if it is a fiducial $\normal$-observer, then $K(t)=K_0/a(t)^2$ with $a(t)$ standing for both curvature and volumetric scale factors along $c(t)$. 
\end{thm}
\proof
This theorem is needless in the flat case, hence we assume $K\ne 0$. Let us take canonical coordinates centered at $c(t_0)=(t_0,0)$ as in theorem \ref{Thm:LocalFormConstantCurvature} with $\phi(t,0)=0$, with no loss of generality. Since $c(t)$ is an integral line of $\normal$, $c(t)=(t,0)$. The condition of being free-falling implies $\grad_{\dot{c}}\dot{c}=\grad_{\normal}\normal=\gradbar\phi=0$ along $c(t)$, that is, $\gradbar\phi(t,0)=0$. In order for the Hubble parameter to be spatially homogeneous up to the first order, we must have 
\begin{equation}
\gradbar_i\hubble(t,0) = - \partial_i\hubble_\psi (t,0)  
= \frac{\partial}{\partial t}\left( \frac{Ka^2\gamma_{ik}x_0^k}{2 \psi(t,0)} \right)= 0 \, ,    
\end{equation}
where we used $\partial_i\hubble_\psi(t,0)= \partial_t (\partial_i\ln\psi)_{x=0}$. Hence, $Ka^2\gamma_{ik}x_0^k=2 \psi(t,0)A_i$, for some constants $A_i$. However, since the coordinates are centered at $(t_0,0)$, $x^i_0(t_0)=0$, that is, $A_i=0$ for every $i=1, \ldots, m-1$. This is possible only if $x_0^i(t)=0$ for any $t$, which proves the first part of the theorem. The rest follows from the fact that, if $c(t)$ is fiducial, then the volume scale factor along it is $a(t)$ so that $K(t)=K_0/a(t)^2$.
\qed

\subsection{Einstein's equations }\label{Sec:EinsteinEquations}

The spatial sections of the intrinsic homogeneous and isotropic spacetimes have constant sectional curvature, which means that the Riemann tensor, the Ricci tensor, and the curvature scalar are, respectively
\begin{equation}
\curvature^{ik}_{j\ell}=K (\delta^i_j\delta^k_\ell-\delta^i_\ell\delta^k_j)
\quad , \quad 
\curvature^{i}_{j}=(m-2)K \delta^i_j
\quad , \textrm{and}\quad 
\curvature= \constant^2 K \, .
\end{equation}  
Using those relations in Einstein's equations $R_\mu^\nu - R/2 \delta_\mu^\nu = T_\mu^\nu$, where we have adopted the convention of inserting the cosmological constant in the energy density and relativistic pressure through the substitutions $\energy \to \energy + \Lambda$ and $\pressure \to \pressure - \Lambda$,  we obtain:
\begin{enumerate}[(i)]
\item Generalized Friedmann equation: 
\begin{equation}\label{Eq:FriedmannEquation}
\frac{1}{2}\constant^2\left( (1-\anisotropy^2)\hubble^2 + K \right) = 
\rho 
\end{equation}
\item Energy flux equation:
\begin{equation}\label{Eq:EnergyFluxEquation}
q_{i} = (m-2) \gradbar_i \,\hubble 
-  \constant \gradbar_k \left(\hubble\anisotropy^k_{i}\right) 
\end{equation}
\item Raychaudhuri's equation:
\begin{eqnarray}\label{Eq:Raychaudhuri}
e^{-\phi}\, \dot{\hubble} + \left(\, 1+ (m-2)\anisotropy^2\right)\,\hubble^2  
&=& 
-  \frac{1}{\constant^2} \left( (m-3) \energy + (m-1)\pressure \right) \nonumber \\
&+& \frac{1}{m-1}\, e^{-\phi}\, \gradbar^2\,e^{\phi} 
\end{eqnarray}
%
%
%
\item Hubble anisotropy equation:
\begin{equation}\label{Eq:HubbleAnisotropyEquation}
e^{-\phi} \frac{\partial}{\partial t}\left(\hubble\anisotropy^k_{i}\right) + (m-1)\hubble^2 \anisotropy^k_{i}  
= \frac{1}{\constant} \left(\pi_i^k + \widehat{\Phi}_i^k\right)  \, ,  
\end{equation}
where
\begin{equation}\label{Eq:TidalTensor}                %
\widehat{\Phi}_i^k = e^{-\phi}\, \gradbar_i\gradbar^{\,k}e^{\phi} - \frac{1}{m-1}\, e^{-\phi}\, \gradbar^2 e^{\phi} \, \delta_i^k \, .
\end{equation}
\end{enumerate}
Note that $\widehat{\Phi}_i^k$ resembles the Newtonian tidal tensor in the weak potential regime $\phi <<1$. For $m=4$, as we make the substitutions $T_{\mu\nu} \to 8\pi G T_{\mu\nu}$, $\hubble \to \theta/3$, $\gradbar_i\phi=\dot{\normal}_i$, and $\anisotropy^k_i \to 3\sigma^k_i/\sqrt{6}\theta$, with the notations $\sigma^2=(\sigma^k_i\sigma^i_k)/2$ and ``$\dot{x}=\grad_\normal x$", we can verify that $\grad_\mu\normal^\mu=e^{-\phi}\, \gradbar^2\,e^{\phi}$, $\grad_\normal\sigma^k_i=e^{-\phi}\partial \sigma^k_i/\partial t$, and the above equations are the vorticity-free ($\omega^\nu=0$) Einstein's equations in Ref. \cite{ellis_mac_marteens},  as in the case of eq. (6.23), for the generalized Friedmann,  eq. (6.20), for the energy flux, eq. (6.4), for Raychaudhuri's, and eq. (6.25), for the Hubble anisotropy equation. 

The intrinsically homogeneous and isotropic spacetimes have been considered earlier in the literature, but not in its most general aspects. There are a lot of quantitative and qualitative results concerning the homogeneous Bianchi models \cite{ExactSolutions, wainwright_ellis_1997}, which encompasses our framework of the free-falling situation ($\gradbar\phi=0$). When $K=0$, it is equivalent to the general Bianchi I class, and when $K \ne 0$, it means we are in special cases of Bianchi V ($K<0$) or Bianchi IX ($K>0$). In the inhomogeneous context ($\gradbar\phi\ne 0$), Einstein's equations for $K = 0$ seem to have been most studied, and a series of exact solutions in different contexts can be found \cite{JMP_1986_WolfThomas, JMP_1986_WolfThomas_2, CQG_2000_Sopuerta,LGGomes_2021_IJMPD, LGGomes_2022_CQG_2}. For the inhomogeneous but isotropic ($\anisotropy=0$) case with general curvature, see Chapter 4 in Ref. \cite{Book_Krasinski_1997} and the references therein.

%
%

\section{The three pieces of the the Cosmological Principle}\label{Sec:H1+H2+H3+EEq}

So far, we have described the three pieces composing the Cosmological Principle. As we put them together two-by-two without the third party, the spacetime cannot be FLRW, even when we admit Einstein's equations to hold. In fact, we have: 
\begin{itemize}
\item H\ref{Hyp:IsotropyHubble} + H\ref{Hyp:IsotropyEMTensor} + Einstein's Equations do not imply H\ref{Hyp:IsotropyCurvature}: In any generalized Robertson-Walker spacetime \cite{GRG_2014_GRWSpacetimes}, obtained by putting $\phi=0$ and $\submetric_{ik}=a(t)^2\submetrict_{ik}(x)$ in formula (\ref{Eq:MetricGeneral}) for the metric $\metric$, every $\normal$-observer is fiducial and shear-free ($\anisotropy=0$). The Hubble parameter is homogeneous ($\hubble=\hubble(t)$) and $\metric$ is a perfect fluid solution of Einstein's equations. However, except when $\submetrict$ has constant curvature, it does not satisfy H\ref{Hyp:IsotropyCurvature}.
\item H\ref{Hyp:IsotropyCurvature} +  H\ref{Hyp:IsotropyEMTensor} + Einstein's Equations do not imply H\ref{Hyp:IsotropyHubble}: Any perfect fluid Bianchi I solution of Einstein's equations with $\anisotropy \ne 0$ can be used as example (see the general barotropic solution in section 4.1 of Ref. \cite{LGGomes_2021_AP}, for instance).
\item H\ref{Hyp:IsotropyCurvature} + H\ref{Hyp:IsotropyHubble} +Einstein's Equations do not imply H\ref{Hyp:IsotropyEMTensor}: for the metric as in corollary \ref{Thm:H1plusH2} with $K=0$, any free-falling $\normal$-observer is fiducial with $\anisotropy=0$. In Ref. \cite{LGGomes_2021_IJMPD} there is such a solution for an imperfect fluid. 
\end{itemize}
However, when we assume all those hypotheses and Einstein's equations to hold, we recover the mathematical formulation of the Cosmological Principle:
\begin{thm}\label{Thm:H1+H2+H3}
An intrinsically homogeneous and isotropic spacetime filled with a perfect fluid where a fiducial observer detects an isotropic Hubble parameter is FLRW around it provided the curvature does not change sign. In this sense,  
\begin{equation}
H\ref{Hyp:IsotropyCurvature} + H\ref{Hyp:IsotropyHubble} + H\ref{Hyp:IsotropyEMTensor} + \textrm{Einstein's equations}
\qquad \Leftrightarrow \qquad \textrm{FLRW.}
\end{equation}
If one of the hypotheses H\ref{Hyp:IsotropyCurvature}, H\ref{Hyp:IsotropyHubble}, or H\ref{Hyp:IsotropyEMTensor} is lacking, the implication cannot hold true. 
\end{thm}
\proof
We use canonical coordinates, as in theorem \ref{Thm:LocalFormConstantCurvature}, centered at any point of the fiducial $\normal$-observer of hypothesis H\ref{Hyp:IsotropyHubble}. According to formula (\ref{Eq:FormsHubble+Anisotropy}),  $\anisotropy=0$ for at least one observer means that $\dot{\submetrict}_{ik}=0$, that is, $\gamma_{ik}(t)=\delta_{ik}$, whith no loss of generality. In particular, $\anisotropy=0$ everywhere. Since the fluid is perfect, the energy flux equation (\ref{Eq:EnergyFluxEquation}) implies $\gradbar\hubble=0$. This condition and the hypothesis H\ref{Hyp:IsotropyHubble} allow us to apply theorem \ref{Thm:WanderingCenterSymmetry}, and therefore obtain $x^i_0(t)=0$ and $K=K_0/a^2$. In particular, we have $\hubble_\psi=0$ and $(\dot{a}/{a})e^\phi\gradbar\phi=\gradbar\hubble=0$, that is, $\phi=0$ with no loss of generality. In short, we have canonical coordinates as in theorem \ref{Thm:LocalFormConstantCurvature}, centered at the fiducial $\normal$-observer of hypothesis H\ref{Hyp:IsotropyHubble}, with $\phi=0$, $x^i_0(t)=0$, $K=K_0/a^2$, and $\gamma_{ik}(t)=\delta_{ik}$. In other words, a FLRW spacetime.    
\qed

The characterization of FLRW spacetimes is an old topic in the literature. The most famous result is the seminal Ehlers-Geren-Sachs theorem \cite{JMP_1968_EGS_Theorem}, stating that in regions where an isotropic CMB is the unique source of energy-momentum, the spacetime must be FLRW. Another very interesting characterization relies upon the redshift against the angular diameter distance up to third order \cite{1999_CQG_Perlick}, which has a beautiful observational appeal. In Geometry, this kind of problem is often referred to as rigidity conditions for the FLRW spacetimes (see \cite{LMP_2023_Avalos_Rigidity} and references therein).

The Stephani-Barnes universes \cite{1973_GRG_Barnes, CMP_1967_Stephani, 1981_GRG_Krasinski, 1983_GRG_Krasinski, Book_Krasinski_1997} are "almost FLRW" spacetimes, characterized by
\begin{equation}
\textrm{Stephani-Barnes} = \left\{ H\ref{Hyp:IsotropyCurvature} + (\anisotropy =0) + H\ref{Hyp:IsotropyEMTensor} + \textrm{Einstein's equations} \right\} \, .
\end{equation}
Here we have unveiled the two subtle characteristics differentiating both, which are the existence of a free-falling $\normal$-observer and the fact that along it the curvature and the volume can be properly scaled. According to the fact that $\gradbar\hubble=0$ in such environments and using theorem \ref{Thm:WanderingCenterSymmetry}, these mean that the "wandering center of symmetry" can be settled as $x^i_0(t)=0$ and that $K\sim a^{-2}$, respectively.

\section{Naturalness for homogeneity on large scales: 
the periodic boundary conditions}
\label{Sec:PeriodicConditions}

\subsection{Periodic boundary conditions}

One distinguished property of the intrinsically homogeneous and isotropic spacetimes is the natural geometric structure they have for building effective homogeneous models. In fact, we can transcribe in our spacetime the concept that the universe, after taking averages on a large scale, is described by an effective model that behaves like an homogeneous one. In such a framework, we can smooth the problems arising in the process of ``averaging" in Cosmology \cite{2011_RPP_Ellis_Clarkson}. This idea of space being divided into a lattice formed by identical regions, with matter behaving equally in each one of its cells, and therefore giving rise to a large-scale homogeneous structure, has appeared in the literature since the 1950s, after the seminal work of Lindquist and Wheeler \cite{Wheeler_1957}, and continued in lines like ``Archipelagian Cosmology" \cite{CliftonFerreira_2009,Clifton_2012,Liu_2015}, ``black hole lattice spacetimes" \cite{bentivegna}, and others\cite{Bruneton_2012,Eingorn2021,Zhuk_2015,Zhuk2021,Hellaby_2012}. Here we extend the ideas first developed in Ref. \cite{LGGomes_2022_CQG_2}.

The notion of tilling space into identical cells is natural to spaces of constant curvature, which can be accomplished in a quite straightforward way. Hence, let us assume that each one of our spatial sections, identified with $\MS$, is one of the following geometric forms: the Euclidean space $\R^{m-1}$, for $K_0=0$, the hyperbolic space $\mathbb{H}^{m-1}$, for $K_0<0$, or the sphere $\mathbb{S}^{m-1}$, for $K_0>0$. Take a discrete subgroup $\Gamma$ of the group of the isometries of the space $\MS$ of constant curvature $K_0$, and assume its action on $\MS$ is free and properly discontinuous so that the quotient $\K=\MS/\Gamma$ is a complete manifold of constant curvature $K_0$ (corollary 2.3.17 in \cite{Wolf}), which we denote by $\K$. For our purposes here, we shall also assume that $\K$ is compact. Hence, we have the following picture: the space $\MS$ is divided into identical regions, called fundamental domains of $\Gamma$ \cite{vinberg}, so that any $\Gamma$-periodic object, like a function invariant by $\Gamma$, for example, turns out to be equivalent to the corresponding object in $\K$. 

Let us translate this scheme to Cosmology by assuming our spacetime to be composed of maximally symmetric spatial sections, that is, one of the standard models above of constant curvature $K_0$. At large scales, the universe is quite homogeneous. That means patterns of galaxy distribution and other physical parameters are distributed virtually in an equal manner along identical regions of typical length $L>>L_0$, with $L_0$ of the order of $100$ Mpc or greater (see Ref. \cite{ObservationalCosmologicalPrinciple} for an up-to-date discussion). These are the cosmological cells, which we identify with the fundamental domains of $\Gamma$, as described above. Therefore, the ``Cosmological Principle in practice" tells us that the inhomogeneities are $\Gamma$-periodically distributed throughout space such that dealing with these local irregularities in each cosmological cell turns out to be equivalent as if we were in the compact manifold $\K$. In order to illustrate it, consider the following two cases:
\begin{itemize}
\item If $K_0=0$ and $m=4$, we could proceed as in Ref.\cite{LGGomes_2022_CQG_2} and take our three-dimensional space $\R^3$ to be divided into an infinite number of cubic boxes of length $L_0$, our cosmological cells. In this case, the discrete subgroup of symmetry is formed by the translations $(x,y,z) \mapsto (x + n_x L_0,y + n_y L_0,z+ n_z L_0)$,  $n_x,n_y,n_z \in \Z$. Hence, the periodicity of galaxy distribution along those boxes would reduce the mathematical approach to the inhomogeneities to the three-dimensional torus $\T^3=\R^3/\Gamma$. In particular, this periodic boundary condition would demand Einstein's equation to be formulated in the torus $\T^3$, the compact manifold that we could identify with any one of the cosmological cells.  
\item If $K_0=-1$, we could first set $m=3$, for the sake of simplicity. Hence our spatial sections would be copies of the hyperbolic plane $H^2$, and the discrete subgroup of symmetries $\Gamma$ preserving orientation, called a Fuchsian group, would lead us to the $g$-connected sum $\K=H^2/\Gamma = \T^2 \# \ldots \# \T^2$,  classified by the genus $g\ge 2$. For instance, if $g=2$, we would formulate our equations in the bi-torus $\T^2 \# \T^2$, which could also be identified with any one of the cosmological cells. In this case, the galactic distribution would resemble one of Escher's ``Circle Limit" masterpieces. For the dimension of interest in Cosmology, $m=4$, the richness of the hyperbolic geometry allows us to decompose the space $H^3$ in an immense amount of possibilities.   
\end{itemize}

The preceding argumentation allows us to formulate the periodic boundary condition by demanding the spatial functions as $\hubble(t,*)$, $\energy(t,*)$, and $\phi(t,*)$ to be $\Gamma$-periodic for all possible time $t$, that is, 
they are such that $f(t,g\cdot x)=f(t,x)$ for all $x \in \MS$, as "$\cdot$" denotes the action of $\Gamma$ on $\MS$ (See \cite{LGGomes_2022_CQG_2} for examples with $K_0=0$). Formally, we have:
\begin{defi}[Periodic Boundary Condition]\label{Def:PeriodicBoudaryCondition}
The spacetime is $\M=\I \times \MS$, $\I \subset \R$ an open interval, and the restriction of its metric $\metric$ to the space section $\{t\}\times \MS$ is $a(t)^2 \gamma(t)$, where $(\MS,\gamma(t))$ is a maximally symmetric Riemannian space of constant curvature $K=K(t)$. Furthermore, for each $t \in \I$, there is a subgroup $\Gamma(t)$ of the isometries of the space $(\MS,\gamma(t))$ such that the quotient $\MS/\Gamma(t)$ endowed with the inherited Riemannian metric induces an isometry onto a compact Riemannian manifold $\K$ with constant curvature $K=K(t)$ and fixed volume $L_0^{m-1}$. Any function or spatial tensor field involved in the dynamics of Einstein's equations is $\Gamma(t)$-periodic in the sense that they are well defined in the quotient $\MS/\Gamma(t)$.
\end{defi}

Following the periodic condition above, we shall often identify any cosmological cell of the tilling imposed by $\Gamma(t)$ at each time $t$ with the compact manifold $\K \cong \MS/\Gamma(t)$. $\K$ does not depend on time, even though $\Gamma(t)$ does. For instance, in the case $m=2$ and $K=0$, the only compact and orientable possibility for the manifold $\K$ is the torus $\T^2$, while $\Gamma(t)$ could be, for instance, the time-dependent subgroup of translations in the Euclidean plane given by $(x,y) \mapsto (x+ e^t L_0 n, y +e^{-3t} L_0  m)$, with $n,m \in \Z$. In this case, we identify any rectangle defined by $0 \le x < e^t L_0$ and $0 \le y < e^{-3t} L_0$ and their $\Gamma(t)$-translations as a cosmological cell, with $\K=\T^2\cong \R^2/\Gamma(t)$, and $\submetrict(t)$ as a flat metric on the torus given by $\submetrict_{11}(t)dx^2+\submetrict_{12}(t)dxdy+\submetrict_{22}(t)dy^2$ for which the canonical infinitesimal area, $\sqrt{\submetrict_{11}\submetrict_{22}-\submetrict_{12}^2}dxdy$, turns out to be $e^{2t}dxdy$, such that $\T^2$ has a constant area $L_0^2$.

\subsection{Naturalness for the averaging process}

The separation of the spatial metric into a scale factor $a(t)$ and a Riemannian metric $\submetrict(t)$, that keeps the constant volume $L_0^{m-1}$ for $\K$, is strategic for the cosmological applications. Note that following the canonical local coordinates of theorem \ref{Thm:LocalFormConstantCurvature} we can put $\submetrict(t)=\submetrict_{ik}(t)dx^idx^k/\psi(t,x)^2$, with $\psi$ given in equation (\ref{Eq:DefPsi}), only if $K(t)=K_0/a(t)^2$. In fact, if we take two instants $t_0$ and $t_1$, with $\gamma_{ik}(t_0)=\delta_{ik}$ and $x_0^i(t_0)=0$, with no loss of generality, and define the affine oriented coordinates change $x^i=x_0^i(t_1) + a^i_ky^k$, with $\gamma_{ik}(t_1)a_j^ia_\ell^k=\delta_{j\ell}$, since $\det(\gamma_{ij})=1$, we get $\det(a^k_i)=1$ as well, that is, $dx^1\wedge \ldots \wedge dx^{m-1}=dy^1\wedge \ldots \wedge dy^{m-1}$. Therefore, the volume of $\K$ at the instant $t_1$ is
\begin{equation}
\frac{V(t_1)}{a(t_1)^{m-1}}
= \int_\K \frac{d^{m-1}x}{\psi(t_1,x)^{m-1}}
=\int_\K \frac{d^{m-1}y}{\left(1 + \frac{K_0}{4}\delta_{ik}y^iy^k \right)^{m-1}}
= \frac{V(t_0)}{a(t_0)^{m-1}} \, .
\end{equation}
Choosing an instant $t_0$ with $a(t_0)=1$ and $V(t_0)=L_0^{m-1}$, we obtain
\begin{equation}
K(t)=\frac{K_0}{a(t)^2} \qquad \Rightarrow \qquad 
V(t) = a(t)^{m-1}\, L_0^{m-1} \, .
\end{equation}
In particular, this occurs if there is a fiducial $\normal$-observer perceiving a Hubble parameter homogeneous to the first order, as in theorem \ref{Thm:WanderingCenterSymmetry}. Moreover, the average value of any $\Gamma(t)$-periodic function does not depend on $a(t)$, for 
\begin{equation}\label{Eq:MeanValue}
\langle F \rangle(t)= \frac{1}{L_0^{m-1}}\int_\K F(t,x) \frac{d^{m-1}x}{\psi(t,x)^{m-1}}\, .
\end{equation}
Using the definitions in equation (\ref{Eq:HubbleFunctions}), and always denoting $\langle\cdot\rangle$ for the mean value along the compact manifold $\K$, we have
\begin{equation}
\frac{d}{dt}\langle F \rangle = \langle \frac{\partial F}{\partial t} \rangle - (m-1)  \langle F\, \hubble_\psi \rangle \, . 
\end{equation}
In the resume, we have a natural geometric structure for the averaging process induced by the discrete subgroup of symmetries of the periodic condition, such that the problem of non-commutativity of the time derivatives is smoothed out, and even disappears if $\hubble_\psi=0$.

\section{Decoupling the geometric and the CMB frames: the need for
intrinsically homogeneous and isotropic spacetimes}
\label{Sec:ReferenceFrame}

Let us now identify where the $\normal$-observers are, the ones to whom the spatial geometry is homogeneous and isotropic. We call it the geometric frame of reference. Besides that, we admit the existence of free-falling, vorticity-free, homogeneous, and isotropic background radiation whose co-moving frame is defined by the time-like vector field $\tilde{\normal}$, with $v=\tilde{\normal}-\normal$ the spatial velocity relative to the $\normal$-observers. The CMB in dimension $m=4$ is given by a perfect fluid energy-momentum tensor with respect to the $\tilde{\normal}$-observers whose pressure has three times the magnitude of the spatially homogeneous energy density  $\widetilde{\energy}_r$, that is,
\begin{equation}
T^{CMB}_{\mu\nu} = \frac{\widetilde{\energy}_r}{3} \left( 4 \, \tilde{\normal}_\mu\tilde{\normal}_\nu + \metric_{\mu\nu}\right) 
\quad , \quad \grad_X\widetilde{\energy}_r=0 \quad \textrm{whenever} \quad \metric(X,\tilde{\normal})=0\, .
\end{equation}
From the $\normal$ viewpoint, this gives an energy-momentum tensor with energy density $\energy_r=\widetilde{\energy}_r$, pressure $\pressure_r=(1+v^2)\energy_r/3$, energy flux $q_r^i = \energy_r v^i$ and stress $\pi_r^{ik}$ given by the trace-less part of $\energy_r v^iv^k$. Note that at any point $p \in \M$ where $v=0$ these different descriptions of the CMB coincide. 

The fact that the CMB is free-falling means that the unique integral curve of $\tilde{\normal}$ passing at $p$ in the instant $t_0$, that is, the curve $c(\tau) \in \M$ such that $c(\tau(t_0))=p$ and $x'(\tau)=\tilde{\normal}(c(\tau))$, is in fact a geodesic. As long as we allow the relative velocity of the CMB frame to vanish along $c(\tau)$, $v(c(\tau))=0$, we realize that $\tau=t$ and $c(t)$ is a free-faling $\normal$-observer. In particular,  $\gradbar\phi(c(t))=0$ for every $t$. In other words, $c(t)$ is a special observer to whom the CMB and constant curvature frames coincide if, and only if, it is placed exactly where the net "gravitational force" vanishes, as in the case of the fiducial observers of definition \ref{Def:FiducialObservers}. 

Due to the considerations above, if we assume that our geometric reference frame does not coincide with the CMB's, except for some fiducial observers symmetrically distributed along the geometric frame, our spacetime is no longer FLRW, but instead, an intrinsically homogeneous and isotropic one. The discrete subgroup of symmetries defining the periodic condition of section \ref{Sec:PeriodicConditions} would be characterized by the disposition of these fiducial $\normal$-observers along space, at least in principle. Our current measurements based on the dipole anisotropy of the CMB can tell us that our peculiar velocity relative to these special observers is something near $300$ km/s in a specific direction \cite{PRL_2021_Quartin}, but cannot distinguish this picture from the canonical FLRW counterpart. Therefore, probing the validity of the hypothesis stating that the CMB and geometric frames do not coincide everywhere could be achieved only through indirect means. For this reason, we must build effective cosmological models over the framework of intrinsically homogeneous and isotropic spacetimes for later scrutinizing their observational validity. In the next section, we comment on how we could start pushing forward this procedure.

\section{Towards an effective homogeneous model from 1/3 of 
the Cosmological Principle}
\label{Sec:EffectiveModel}

Among the intrinsically homogeneous and isotropic spacetimes, those satisfying the periodic boundary condition of section \ref{Sec:PeriodicConditions} have a natural structure to build average values of the physical quantities. In particular, we can mount effective cosmological models similar to the standard $\Lambda$CDM one, but still quite different due to the characteristics of the geometric background. However, there is a warning in advance to the reader: our intention here is to furnish a glimpse of the potential such models can have in Cosmology. A complete development of such a scheme will appear soon elsewhere.     

We start by setting the dimension to be $m=4$ and the scale of homogeneity as $L_0$. We assume the existence of a set of fiducial observers symmetrically spread in space based on whom we define the geometric frame and the discrete subgroup of symmetries, $\Gamma(t)$. Hence, they define the periodic boundary condition set in section \ref{Sec:PeriodicConditions}, and thus the typical cosmological cell denoted by $\K$ and identified with a compact manifold endowed with a time-dependent family of metrics $\submetric(t)=a(t)^2\submetrict(t)$. In canonical coordinates, they are represented as in theorem \ref{Thm:WanderingCenterSymmetry}, with constant curvature $K(t)=K_0/a(t)^{2}$  such that $\K$ has volume $a(t)^3L_0^3$. We define the mean value of any continuous function as in the formula (\ref{Eq:MeanValue}). We calibrate the $\normal$-observers' clocks to show the same time $t_0$ for "today", that is, redshift $z=0$, when we set $a(t_0)=1$. Our fundamental assumption is that those fiducial observers are simultaneously at rest in the CMB and geometric frames.

The energy density is split into two parts. The first one emulates the homogeneous $\Lambda$CDM model. For this, we borrow the notation from the standard picture and denote the critical density by $\energy_c$, which we are going to use as a reference for the energy density magnitude. As usual, we define the density parameters for the dark energy, radiation, and matter by, respectively, \cite{Weinberg}
\begin{equation}
\energy_c=3 \langle \hubble \rangle^{2}(t_0) \quad ,\quad
\Omega_\Lambda=\frac{\Lambda}{\energy_c} \quad ,\quad 
\Omega_r=\frac{\langle \energy_r \rangle(t_0)}{\energy_c} \quad ,\quad 
\Omega_m=\frac{\langle \energy_m \rangle(t_0)}{\energy_c} \, , 
\end{equation}
where $\Lambda$ is the cosmological constant, $\energy_r(t,x)$ is the energy density due to radiation, and $\energy_m(t,x)$ the part coming from baryonic and dark matter inside the cosmological cells. The first one is composed mainly of free-falling background radiation whose frame is moving with the spatial velocity $v$ described in section \ref{Sec:ReferenceFrame}. The total energy density becomes
\begin{equation}
\frac{\energy(t,x)}{\energy_c}
= \Omega_\Lambda + \frac{\Omega_m}{a^{3}}+ \frac{\Omega_{r}}{a^4}  
+ \energyint(t,x) \, .
\end{equation}
The last term in the equation above is the internal energy density (in units of $\energy_c$), which should be understood as the contribution of anything that could make the energy density depart from the $\Lambda$CDM proposal. Its physical interpretation would vary with the background assumptions on the matter/radiation content, which could stand for things as the internal motion and interactions within each cosmological cell, the energy of some fields inside them, corrections due to the fact that the geometric and CMB's frames are not the same, and so on. In general, the real aspect of it will only show up when we further investigate Einstein's equations in a desired given situation. The density parameter associated with it is defined as 
\begin{equation}
\omegap_\mu(a) = \langle\energyint\rangle    
\quad , \quad \omegap_\mu(1):=\Omega_\mu \, .
\end{equation}
Its dependence on the scale factor, interpretation, and relevance to the cosmological picture must be analyzed case by case, which should vary with different configurations for the energy-momentum tensor.

The inhomogeneities inside each cosmological cell $\K$ also contribute with a density parameter, the one representing the standard deviation of the true Hubble parameter $\hubble$ from its homogeneous substitute $\langle\hubble\rangle$. Formally, it is defined as  
\begin{equation}
\omegap_I(a)\, \energy_c = \langle\hubble^2\rangle - \langle\hubble\rangle^2 
= \langle\left(\hubble - \langle\hubble\rangle\right)^2\rangle 
\quad , \quad \omegap_I(1):=\Omega_I \, .
\end{equation}
Hence, $\omegap_I(a(t))=0$ if, and only if, the Hubble parameter is homogeneous at time $t$, that is, $\hubble(t,\cdot)=\langle\hubble\rangle(t)$. It is important to note here that, while the cosmic expansion occurs apparently in a homogeneous way driven by $a(t)$, its perception through the $\normal$-observers must take into account their different proper times, $d\tau=e^{\phi}dt=da/\hubble$, since the geometric frame is not free-falling. Hence, $\omegap_I(a)$ would also be a measure of the dispersion of the $\normal$-proper times along space.

The average value of the standard deviation of the anisotropy along the cosmological cells gives us another density parameter, which is defined as 
\begin{equation}
\omegap_A(a) = \frac{\langle\hubbleA^2\rangle}{\energy_c}
\quad , \quad \omegap_A(1):=\Omega_A \, ,
\end{equation}   
where $\hubbleA$ has been introduced in equation (\ref{Eq:AnisotropyStandardDeviation}). Its magnitude has both local and global contributions. The first one would come from the irregular distribution of matter inside each cosmological cell, even for vacuum, since the Newtonian-like tidal term $\gradbar^i\gradbar_k\phi$ would behave as a source for the anisotropy equation (\ref{Eq:HubbleAnisotropyEquation}). On the other hand, the global counterpart would be characteristic of a large-scale effect. In that case, it seems reasonable to expect those contributions to be quite similar to their counterpart in the homogeneous Bianchi models. It is common to attribute to this Hubble anisotropy the behavior $\Omega_A/a^6$, which is characteristic of a flat and perfect fluid Bianchi I universe \cite{PRD_2019_BianchiILCDM_Tedesco}. Notwithstanding, the most expected situation is quite different, for the FLRW universes are in general unstable equilibrium points in the Kasner disc \cite{CollinsHawking1973}, and there is a plethora of possibilities for the asymptotic behavior even in the simplest case of vanishing curvature \cite{Misner68,Calogero2008,LeBlanc_1997,LGGomes_2017_CQG,LGGomes_2021_IJMPD,LGGomes_2023_EPJC_1}. Furthermore, kind of behaviors such as the BKL picture of the early universe are not excluded here, at least not a priori, for they would be a possibility in the positive curvature case \cite{PRL_1969_Mixmaster_Misner, AdvP_1970_BKL, Book_2014_Berger_BKL}. 

As we put all the contributions from the internal energy density, inhomogeneities, and anisotropies, we recover an average version of the Friedmann equation (\ref{Eq:FriedmannEquation}), which turns out to be 
\begin{equation}\label{Eq:FriedmannAveraged}
3 \frac{\langle\hubble\rangle^2}{\energy_c}
= \Omega_\Lambda + \frac{\Omega_{r}}{a^4} + \frac{\Omega_m}{a^{3}} + \frac{\Omega_{K}}{a^2} +
\omegap_\mu(a) + \frac{15}{4}\omegap_A(a) - 3\omegap_I(a)\, ,
\end{equation}
where $\Omega_K=-\,3 K_0/\energy_c$ is the density parameter for the curvature, as usual. Clearly, depending on the values of the constants $\Omega_\mu$, $\Omega_I$, and $\Omega_A$, a new interpretation of the constituents of the Universe might emerge. In particular, $\energy_c$ is no longer a ``critical" value that determines the sign of the curvature, for
\begin{equation}
\langle\energy\rangle(t_0)-\energy_c = 3\Omega_I-\frac{15}{4}\Omega_A-\Omega_K \, .
\end{equation}
The $\Lambda$CDM model is recovered as we set $\Omega_\mu \approx \Omega_A \approx \Omega_I \approx 0$. Nonetheless, the vanishing of these new constants does not imply the background geometry is FLRW. Instead, it imposes some constraints on the average values of these inhomogeneous parameters. That is, even the $\Lambda$CDM model might surge from an inhomogeneous background in this approach. 

As we remarked before, we shall not extend the development of this effective averaged model, for it requires much more effort and would take us out of the scope of this manuscript. However, the arguments put above show us that it is possible to have an environment where a $\Lambda$CDM-like model naturally coexists with non-linear inhomogeneities and anisotropies, and therefore turns out to be an interesting arena to investigate how one affects the others in different phases of the universe. Hence, it is natural to ask here whether a highly isotropic CMB would imply $\Omega_A \approx 0$ or how intense would be the natural Hubble tension of these models. We finish by commenting on these two topics below.

Since the CMB radiation seems to be very isotropic as it is seen by the fiducial observers to whom the geometric and CMB frames coincide, as mentioned in section \ref{Sec:ReferenceFrame}, there is a highly spread common sense that this fact must imply $\Omega_A$ to vanish, and observational analysis very often take this implication for granted \cite{PRL_2016_Saadeh_Isotropy}. Furthermore, the "cosmic no-hair" theorem \cite{Wald1983}, which excludes the positive curvature situation, states that a positive cosmological constant would exponentially kill any anisotropy in the universe. However, recent probes suggest that the Hubble sky might not be isotropic on cosmic scales \cite{Anisotropy01, Anisotropy02, ObservationalCosmologicalPrinciple} and the no-hair argument does not necessarily imply a collapse of the Hubble anisotropy in the sky we observe today, for the time required for that mechanism to start is greater than the age of our universe (see sec. 6 in Ref.\cite{LGGomes_2023}). In order to show how we can shed some light on this conundrum, let us permit ourselves to interpret $\phi$ as a Newtonian gravitational potential so that CMB/fiducial $\normal$-observers are in preferred locations of the late-time universe where the net gravitational force on them vanishes, as for instance, in the center of mass of their local galactic cluster. In this context, it is not clear the constraints a small anisotropy of the CMB would impinge in the tidal tensor $\Phi^i_k$ nor in the Hubble anisotropy, neither it seems clear the dynamical behavior they would develop through the equation (\ref{Eq:HubbleAnisotropyEquation}). Even if $\anisotropy$ is small, the simple fact that they could have different orders of magnitude would leave observable marks in the deceleration parameter \cite{LGGomes_2022_CQG_1}. Therefore, our background spacetime provides a natural theoretical arena for further investigations on the behavior of those anisotropies and how they would leave their observational signals in the late-time sky. 

We finally call attention to the fact that a natural Hubble tension must arise in our framework in the late-time universe. The measurement of the Hubble constant from the CMB analysis would presumably come together with their cosmological partners from the $\Lambda$CDM model, since they are meaningful in the time of last scattering, as it is very often considered \cite{AA_2018_Planck_CosmParam}. However, there is no reason to believe, at least a priory, that we could keep the same interpretation for the new parameters composing the average Friedmann equation (\ref{Eq:FriedmannAveraged}). In fact, it is the other way around. For the sake of simplicity, let us consider only the isotropic context, that is, assume $\omegap_A=0$, which implies $\hubble_\psi=0$ (see theorem \ref{Thm:WanderingCenterSymmetry}). Under this condition, for the flat spatial sections, it has been verified that the density contrast in our model is compatible with our expectations: it allows a very homogeneous early state to become more and more inhomogeneous as time goes by in the expansion scenario \cite{LGGomes_2022_CQG_2}. Therefore, if we think that from the time of the last scattering through the era where the large-scale structures were formed and until today, $\omegap_I$ might have significantly changed. Hence, it seems improbable to conceive it in the determination of the Hubble constant from the CMB measurements, even knowing that some effects due to the late-time inhomogeneities leave their imprints there, as the integrated Sachs-Wolf\cite{Weinberg}, for instance. On the other hand, the behavior of $\omegap_\mu$ is more difficult to assert in this general context, but it might as well give rise to some new parameters invisible to the CMB and/or some possible contributions to the $\Lambda$CDM existing ones. In the surroundings of the observers to whom the geometric and CMB frames coincide the universe is virtually FLRW, for $\gradbar\phi=0$ along their paths. Hence, the redshift measurements and the luminosity distance on large scales $L>>L_0$ must indeed be dealt with as in the homogeneous and isotropic standard model, except for the fact that we use the mean Hubble parameter in the averaged Friedmann equation (\ref{Eq:FriedmannAveraged}). From this, we readily conclude that the usual FLRW formula for the luminosity distance would be a good approximation, that is, the formula  (Sec. 7.4.5 in Ref. \cite{ellis_mac_marteens})
\begin{equation}
d_L(z) = (1+z)\, S_{K_0}\left( \int_0^z \frac{dz'}{\langle\hubble\rangle(z')} \right)
\quad , \quad 
1+z = \frac{1}{a} \, , 
\end{equation}
with $S_{K_0}(x)=\sin x$, for $K_0 >0$, $S_{K_0}(x)=\sinh x$, for $K_0 <0$, and $S_{K_0}(x)=x$, for $K_0 =0$. Therefore, the discrepancy in the determination of the Hubble constant from the CMB and from the Luminous distance observations is inexorably present in our effective model. 

In the literature, one can find different tentatives to solve the Hubble tension through the inhomogeneous and anisotropic context (see section 3.1. of Ref. \cite{CQG_2021_Hubble_Tension_Review} and the references therein). Whether the scheme above will shed some light on or, on the contrary, intensify the empiric Hubble tension in Cosmology, is a matter of further analysis, that can only be known as we probe inside the parameters $\omegap_I$ and $\omegap_\mu$ through Einstein's equations in specific situations and examine them against real observations. Regardless of the outcome of such investigations, all the general considerations we have made so far are in order to show that the intrinsically homogeneous and isotropic spacetimes are worth investigating, and the effective $\Lambda$CDM-like models arising naturally from them can shed light on some of the most elusive problems encountered in Cosmology today.

\section{Final remarks}
\label{Sec:FinalRemaks}

In this manuscript, we have investigated the foundations of the Cosmological Principle, which has been broken into three parts. The first, mathematical, ensures the homogeneity and isotropy of the spatial geometry to describe the universe, thus defining the geometric frame. The second, observational, guarantees isotropic expansion along fiducial observers. The third, physical, assumes a perfect fluid form for the overall combination of matter and radiation. As we permit ourselves to decouple the CMB from the geometric frame, we remain with the more general structure where just the first of those pieces is kept, the so-called intrinsically homogeneous and isotropic spacetimes. In this context, we have an interesting geometric framework where taking spatial averages and mounting $\Lambda$CDM-like effective models, with a Hubble tension inexorably attached, are naturally conceived. All these properties make them a prolific arena for further theoretical investigations and astrophysical applications.

\section*{Acknowledgments}

The author is thankful for the support from FAPEMIG, project number RED-00133-21.

\section{References}

\bibliography{FoundationsCP.bib}

\providecommand{\newblock}{}
\begin{thebibliography}{10}
\expandafter\ifx\csname url\endcsname\relax
  \def\url#1{{\tt #1}}\fi
\expandafter\ifx\csname urlprefix\endcsname\relax\def\urlprefix{URL }\fi
\providecommand{\eprint}[2][]{\url{#2}}

\bibitem{Peebles}
Peebles P~J~E 1993 {\em Principles of Physical Cosmology\/} (Princeton
  University Press)

\bibitem{Weinberg}
Weinberg S 2008 {\em {Cosmology}\/} (Oxford University Press) ISBN
  978-0-19-852682-7

\bibitem{ABDALLA202249}
Abdalla E {\em et~al.\/} 2022 {\em Journal of High Energy Astrophysics\/} {\bf
  34} 49--211 ISSN 2214-4048

\bibitem{ObservationalCosmologicalPrinciple}
Aluri P~K {\em et~al.\/} 2023 {\em Class. Quant. Grav.\/} {\bf 40} 094001
  (\textit{Preprint} \eprint{2207.05765})

\bibitem{Anisotropy01}
Fosalba P and Gaztañaga E 2021 {\em Monthly Notices of the Royal Astronomical
  Society\/} {\bf 504} 5840--5862 ISSN 0035-8711 (\textit{Preprint}
  \eprint{https://academic.oup.com/mnras/article-pdf/504/4/5840/38104018/stab1193.pdf})
  \urlprefix\url{https://doi.org/10.1093/mnras/stab1193}

\bibitem{Anisotropy02}
Yeung S and Chu M~C 2022 {\em Phys. Rev. D\/} {\bf 105} 083508
  (\textit{Preprint} \eprint{2201.03799})

\bibitem{Wald}
Wald R~M 1984 {\em General Relativity\/} (The University of Chicago Press)

\bibitem{Wolf}
Wolf J~A 2011 {\em Spaces of Constant Curvature\/} (AMS Chelsea Publishing)

\bibitem{vinberg}
Vinberg E 1993 {\em Geometry II\/} (Springer, Berlin, Heidelberg)

\bibitem{ellis_mac_marteens}
Ellis G~F~R, Maartens R and MacCallum M~A~H 2012 {\em Relativistic Cosmology\/}
  (Cambridge University Press)

\bibitem{coll_79}
Collins C~B and Szafron D~A 1979 {\em Journal of Mathematical Physics\/} {\bf
  20} 2347

\bibitem{coll1_79}
Szafron D~A and Collins C~B 1979 {\em Journal of Mathematical Physics\/} {\bf
  20} 2354

\bibitem{coll2_79}
Collins C~B and Szafron D~A 1979 {\em Journal of Mathematical Physics\/} {\bf
  20} 2362

\bibitem{Hwk_Ellis}
Ellis G~F~R and Hawking S~W 1975 {\em The Large Scale Structure of
  Space-Time\/} (Cambridge University Press)

\bibitem{MTW}
Misner C, Thorne K~S and Wheeler J~A 2017 {\em Gravitation\/} (Princeton
  University Press)

\bibitem{LGGomes_2023}
L~G~Gomes M A C~Nogueira L~R~d~S 2023 {\em Submitted\/}

\bibitem{LGGomes_2021_IJMPD}
Bittencourt E, Gomes L and Santos G 2021 {\em International Journal of Modern
  Physics D\/} {\bf 30} 2150033

\bibitem{LGGomes_2022_CQG_2}
Bittencourt E, Gomes L~G and Santos G~B 2022 {\em Classical and Quantum
  Gravity\/} {\bf 39} 225008
  \urlprefix\url{https://dx.doi.org/10.1088/1361-6382/ac96c3}

\bibitem{1973_GRG_Barnes}
Barnes A 1973 {\em General Relativity and Gravitation\/} {\bf 4} 105--129

\bibitem{CMP_1967_Stephani}
Stephani H 1967 {\em Communications in Mathematical Physics\/} {\bf 4} 137--142

\bibitem{1981_GRG_Krasinski}
Krasinski A 1981 {\em General Relativity and Gravitation\/} {\bf 13} 1021--1035

\bibitem{1983_GRG_Krasinski}
Krasi{\'n}ski A 1983 {\em General relativity and gravitation\/} {\bf 15}
  673--689

\bibitem{Book_Krasinski_1997}
{Krasinski} A 1997 {\em Inhomogeneous Cosmological Models\/} (Cambridge
  University Press)

\bibitem{ExactSolutions}
Stephani H, Kramer D, MacCallum M, Hoenselaers C and Herlt E 2003 {\em Exact
  Solutions of Einstein's Field Equations\/} 2nd ed Cambridge Monographs on
  Mathematical Physics (Cambridge University Press)

\bibitem{wainwright_ellis_1997}
Wainwright J and Ellis G~F~R 1997 {\em Dynamical Systems in Cosmology\/}
  (Cambridge University Press)

\bibitem{JMP_1986_WolfThomas}
Wolf T 1986 {\em Journal of Mathematical Physics\/} {\bf 27} 2340--2353

\bibitem{JMP_1986_WolfThomas_2}
Wolf T 1986 {\em Journal of mathematical physics\/} {\bf 27} 2354--2359

\bibitem{CQG_2000_Sopuerta}
Sopuerta C~F 2000 {\em Classical and Quantum Gravity\/} {\bf 17} 4685--4695

\bibitem{GRG_2014_GRWSpacetimes}
Chen B~Y 2014 {\em General Relativity and Gravitation\/} {\bf 46} 1--5

\bibitem{LGGomes_2021_AP}
Bizarria B~B, Silva G~A~S, Gomes L~G and Clavijo W~O 2021 {\em Annals of
  Physics\/} {\bf 432} 168571 ISSN 0003-4916
  \urlprefix\url{https://www.sciencedirect.com/science/article/pii/S0003491621001779}

\bibitem{JMP_1968_EGS_Theorem}
Ehlers J, Geren P and Sachs R~K 1968 {\em Journal of Mathematical Physics\/}
  {\bf 9} 1344--1349

\bibitem{1999_CQG_Perlick}
Hasse W and Perlick V 1999 {\em Classical and Quantum Gravity\/} {\bf 16} 2559

\bibitem{LMP_2023_Avalos_Rigidity}
Avalos R 2023 {\em Letters in Mathematical Physics\/} {\bf 113} 98

\bibitem{2011_RPP_Ellis_Clarkson}
Clarkson C, Ellis G, Larena J and Umeh O 2011 {\em Reports on Progress in
  Physics\/} {\bf 74} 112901

\bibitem{Wheeler_1957}
Lindquist R and Wheeler J 1957 {\em Reviews of Modern Physics\/} {\bf 29}
  432--443

\bibitem{CliftonFerreira_2009}
Clifton T and Ferreira P~G 2009 {\em Phys. Rev. D\/} {\bf 80} 103503

\bibitem{Clifton_2012}
Clifton T, Ferreira P~G and O'Donnell K 2012 {\em Phys. Rev. D\/} {\bf 85}(2)
  023502

\bibitem{Liu_2015}
Liu R 2015 {\em Physical Review D\/} {\bf 92} 063529

\bibitem{bentivegna}
Bentivegna E, Clifton T, Durk J, Korzy{\'{n}}ski M and Rosquist K 2018 {\em
  Classical and Quantum Gravity\/} {\bf 35} 175004

\bibitem{Bruneton_2012}
Bruneton J~P and Larena J 2012 {\em Classical and Quantum Gravity\/} {\bf 29}
  155001

\bibitem{Eingorn2021}
Eingorn M, Canay E, Metcalf J~M, Brilenkov M and Zhuk A 2021 {\em Universe\/}
  {\bf 7} 469

\bibitem{Zhuk_2015}
Brilenkov M, Eingorn M and Zhuk A 2015 {\em The European Physical Journal C\/}
  {\bf 75} 217(1--10)

\bibitem{Zhuk2021}
Eingorn M, McLaughlin A, Canay E, Brilenkov M and Zhuk A 2021 {\em Universe\/}
  {\bf 7} 101

\bibitem{Hellaby_2012}
Hellaby C 2012 {\em Journal of Cosmology and Astroparticle Physics\/} {\bf
  2012} 043--043

\bibitem{PRL_2021_Quartin}
Ferreira P~d~S and Quartin M 2021 {\em Phys. Rev. Lett.\/} {\bf 127}(10) 101301
  \urlprefix\url{https://link.aps.org/doi/10.1103/PhysRevLett.127.101301}

\bibitem{PRD_2019_BianchiILCDM_Tedesco}
Akarsu {\"O}, Kumar S, Sharma S and Tedesco L 2019 {\em Physical Review D\/}
  {\bf 100} 023532

\bibitem{CollinsHawking1973}
{Collins} C~B and {Hawking} S~W 1973 {\em The Astrophysical journal\/} {\bf
  180} 317--334

\bibitem{Misner68}
{Misner} C~W 1968 {\em The Astrophysical Journal\/} {\bf 151} 431

\bibitem{Calogero2008}
Calogero S and Heinzle J~M 2009 {\em Annales Henri Poincare\/} {\bf 10}
  225--274 (\textit{Preprint} \eprint{0809.1008})

\bibitem{LeBlanc_1997}
LeBlanc V~G 1997 {\em Classical and Quantum Gravity\/} {\bf 14} 2281--2301
  \urlprefix\url{https://doi.org/10.1088/0264-9381/14/8/025}

\bibitem{LGGomes_2017_CQG}
Bittencourt E, Gomes L~G and Klippert R 2017 {\em Classical and Quantum
  Gravity\/} {\bf 34} 045010
  \urlprefix\url{https://dx.doi.org/10.1088/1361-6382/aa5994}

\bibitem{LGGomes_2023_EPJC_1}
Dias F~S, Gomes L~G and Mello L~F 2023 {\em The European Physical Journal C\/}
  {\bf 83} \urlprefix\url{https://doi.org/10.1140/epjc/s10052-023-11364-7}

\bibitem{PRL_1969_Mixmaster_Misner}
Misner C~W 1969 {\em Physical Review Letters\/} {\bf 22} 1071

\bibitem{AdvP_1970_BKL}
Belinskii V~A, Khalatnikov I~M and Lifshitz E~M 1970 {\em Advances in
  Physics\/} {\bf 19} 525--573

\bibitem{Book_2014_Berger_BKL}
Berger B~K 2014 Singularities in cosmological spacetimes {\em Springer Handbook
  of Spacetime\/} (Springer) pp 437--460

\bibitem{PRL_2016_Saadeh_Isotropy}
Saadeh D, Feeney S~M, Pontzen A, Peiris H~V and McEwen J~D 2016 {\em Physical
  review letters\/} {\bf 117} 131302

\bibitem{Wald1983}
Wald R~M 1983 {\em Phys. Rev. D\/} {\bf 28} 2118--2120

\bibitem{LGGomes_2022_CQG_1}
Gomes L~G 2022 {\em Class. Quant. Grav.\/} {\bf 39} 027001

\bibitem{AA_2018_Planck_CosmParam}
Aghanim N, Akrami Y, Ashdown M, Aumont J, Baccigalupi C, Ballardini M, Banday
  A, Barreiro R, Bartolo N, Basak S {\em et~al.\/} 2020 {\em Astronomy \&
  Astrophysics\/} {\bf 641} A6

\bibitem{CQG_2021_Hubble_Tension_Review}
Di~Valentino E, Mena O, Pan S, Visinelli L, Yang W, Melchiorri A, Mota D~F,
  Riess A~G and Silk J 2021 {\em Classical and Quantum Gravity\/} {\bf 38}
  153001

\end{thebibliography}

\end{document}